\documentclass[conference, 10pt, review]{IEEEtran}
\IEEEoverridecommandlockouts

\usepackage{cite}
\usepackage{amsmath,amssymb,amsfonts}
\usepackage{algorithmic}
\usepackage{graphicx}
\usepackage{textcomp}
\usepackage{makecell}
\usepackage{algorithm}
\usepackage{booktabs}
\usepackage{multirow}
\usepackage{svg}
\usepackage[table]{xcolor}
\usepackage{booktabs}
\usepackage{tabularx}
\usepackage{pifont}
\usepackage{xcolor}
\usepackage{adjustbox} 
\usepackage[colorlinks,linkcolor=purple,citecolor=purple]{hyperref}

\begin{document}

\title{Unfolding an Atomistic World: Atomistic Simulation of Reactor Pressure Vessel Steel Across Year-and-Meter Scales}



\author{
Haozhi Han\textsuperscript{1,2}\IEEEauthorrefmark{1}
Ruge Zhang\textsuperscript{1,3}\IEEEauthorrefmark{1},
Haoquan Chen\textsuperscript{1,4}\IEEEauthorrefmark{1},
Yifeng Chen\textsuperscript{2}\IEEEauthorrefmark{2},
Haipeng Jia\textsuperscript{3},\\
Liang Yuan\textsuperscript{3},
Yunquan Zhang\textsuperscript{3}\IEEEauthorrefmark{2},
Ting Cao\textsuperscript{1},
Yunxin Liu\textsuperscript{1},
Ya-Qin Zhang\textsuperscript{1},
and Kun Li\textsuperscript{1}\IEEEauthorrefmark{2}\\[6pt]

\textsuperscript{1}Tsinghua University, Beijing, China\\
\textsuperscript{2}School of Computer Science, Peking University, Beijing, China\\
\textsuperscript{3}Institute of Computing Technology, Chinese Academy of Sciences, Beijing, China\\
\textsuperscript{4}Sun Yat-sen University, Guangzhou, China\\
\textsuperscript{*}Equal contribution
\textsuperscript{†}Corresponding authors (likun@air.tsinghua.edu.cn, zyq@ict.ac.cn, cyf@pku.edu.cn)\\


}


\maketitle

\begin{abstract}

Lifetime prediction of reactor pressure vessel (RPV) steel requires bridging atomistic degradation mechanisms with service-scale spatial and temporal regimes, from ångströms and picoseconds to meters and decades. 
Existing engineering-scale models provide long-range reach but rely on fitted degradation laws, while recent atomistic kinetic Monte Carlo (AKMC) advances still fail to achieve year-and-meter scales coverage. 
We present \textit{AtomWorld}, an atomistic world-modeling framework for RPV steel lifetime simulation co-designed with leadership-scale supercomputing through three tightly coupled layers:
(1) \textit{algorithm}: \textit{AtomWorld} recasts classical AKMC as an atomistic world model that learns consequence-aware state transitions over the \textit{ab initio} energy landscape. 
(2) \textit{HPC}: it co-designs this formulation with modern supercomputers yielding a compute-dense, synchronization-light, and communication-efficient execution pipeline.
(3) \textit{application}: it extends atomistic world-modeling to engineering-scale simulation through a physically grounded voxel-parallel framework, offering a scalable pathway from local atomistic dynamics to engineering-scale degradation evolution.
We demonstrate a paradigm shift in atomistic simulation: 
\textit{AtomWorld} achieves 161.9$\times$ to 452.3$\times$ speedup over classical AKMC for advancing one second of physical time on a single NVIDIA A100 GPU, with the advantage increasing monotonically as lattice size grows.
These capabilities are sustained across 4 leadership supercomputers with 92–97\% scaling efficiency.

\end{abstract}



\section{Overview of the Problem}

\begin{figure*}
    \centering
    \includegraphics[width=1\linewidth]{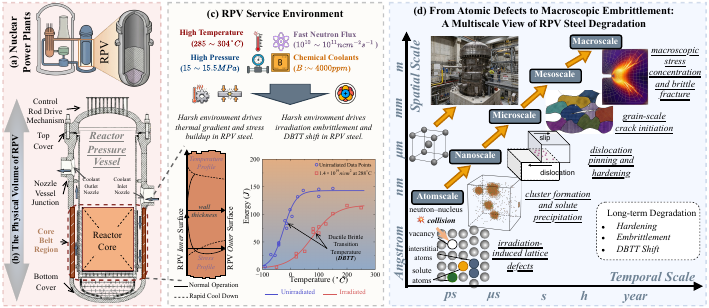}
    \caption{
    Problem context of RPV steel degradation: full-scale structural setting, harsh in-service thermo-irradiation conditions, and the multiscale spatial–temporal evolution from irradiation-induced atomic defects to macroscopic embrittlement.
    (a) and (b) place the RPV in its engineering context and mark the physically relevant vessel regions and dimensions. (c) summarizes the coupled service environment—temperature, pressure, neutron irradiation, and coolant chemistry—that drives stress buildup and irradiation-induced degradation in RPV steels. (d) depicts the cross-scale degradation pathway, in which atomic defects evolve into clusters, precipitation, dislocation pinning, crack initiation, and ultimately hardening, embrittlement, and DBTT shift.
    }
    \label{fig:overall}
\end{figure*}

Predicting the lifetime evolution of nuclear materials under extreme irradiation and thermal environments remains a fundamental challenge in nuclear energy systems~\cite{KOLLURI2023112236, ZINKLE2013735, Little01052006, 10.3389/fnuen.2023.1253974}.
Reactor pressure vessel (RPV) steel is one of its most consequential cases, as it is the irreplaceable safety-critical structural material of the reactor, and its degradation directly constrains RPV lifetime and safe operation~\cite{ma15248761, Mukhtar_Khattak_Rafique_Zareen_2020, TACHIBANA2004103, TIMOFEEV2004703, ALMAZOUZI20113403}.
As illustrated in Fig.~\ref{fig:overall},  decades of irradiation~\cite{SONG201933, KULESHOVA2017247, BECQUART2005111}, thermal aging~\cite{STYMAN201286, FUKAKURA1993423}, and mechanical loading~\cite{TIMOFEEV2004703, KOLLURI2017138} progressively induce irreversible embrittlement and strength loss, pushing the material toward its safety limits. 
The central difficulty lies in an extreme scale gap: the governing degradation mechanisms originate from atomistic events, yet the evolution and failure that matter in practice unfold over service-scale temporal and spatial regimes, spanning picoseconds to years and from ångströms to meters~\cite{MILLER2007145, zelenty2015understanding, ZEMAN2007272, KULESHOVA2020152362}. 

Historically, this challenge has been addressed through engineering-scale models that deliver service-scale temporal and spatial reach~\cite{Odette01061998, KWON20031549, gokhman2008, KE2022153910, 10.1017/S1431927617000162, Odette_Wirth_Bacon_Ghoniem_2001, LIN2024520, CHEIMARIOS20105018, DONG2025124911},  covering RPV-relevant regimes of hundreds of millimeters in wall thickness, more than 10 meters in axial height, and 40-60 years of RPV lifetime~\cite{Mukhtar_Khattak_Rafique_Zareen_2020, JUMEL2005125, TANAKA201526, knott2013structural}.
While useful for design and lifetime assessment, such approaches rely on constitutive closures and fitted degradation rules that become difficult to derive and validate when long-term material behavior is ultimately governed by atomistic events at the ångström and picosecond scales under harsh reactor-relevant conditions~\cite{XIE2015196, osti_1111015, osti_1473611}, including temperatures of $285$--$304^\circ\mathrm{C}$, high pressures of $15$--$15.5\mathrm{MPa}$, fast-neutron fluxes of $10^{10}$--$10^{11}\mathrm{ncm^{-2}s^{-1}}$, and a corrosive coolant environment~\cite{article1111, jne6040048, doi:10.1504/IJNKM.2010.037072}.
They therefore achieve service-scale reach only by sacrificing atomistic fidelity to the mechanisms that ultimately govern material failure~\cite{osti_1473611, osti_1369375}.

At the opposite extreme, the HPC community has increasingly sought to recover atomistic fidelity by scaling atomistic kinetic Monte Carlo (AKMC)~\cite{doi:10.1137/120889459, 10.1145/2966884.2966908, 8421558, 10.1145/3295500.3356165, 10.1145/3458817.3476174, 10.1109/SC41406.2024.00097, 10.1145/3712285.3759781}, an event-driven simulation method for material degradation~\cite{Xu_2012, KIM2022111620, SOISSON201055}, from billion-atom~\cite{10.1145/3295500.3356165} to quadrillion-atom~\cite{10.1145/3712285.3759781} systems through machine-learning-assisted energetic modeling~\cite{10.1109/SC41406.2024.00097} and large-scale supercomputing~\cite{10.1145/3295500.3356165, 10.1145/3712285.3759781}.
Yet scaling system size alone does not deliver the temporal and spatial reach required for RPV lifetime prediction. 
In time, as the system grows, more candidate events and higher total transition rates reduce the physical time advanced per step, while low-barrier, near-reversible local transitions induce severe super-basin trapping over long horizons~\cite{10.1145/3295500.3356165, jassar2025challenges}. 
As a result, {even state-of-the-art methods would still require approximately 30 years of wall-clock time} on a leading supercomputer to simulate a single year of RPV material evolution~\cite{xu2023redesigning, 10.1145/3295500.3356165}. 
In space, even record-scale atomistic simulations still cover only on the order of $10^5 \mu m^3$ of material, far from the macroscopic domains required for RPV applications~\cite{10.1145/3712285.3759781}. 
Conventional AKMC therefore remains unable to simultaneously achieve atomistic fidelity, lifetime-scale temporal advancement, and engineering-scale spatial coverage.

This limitation raises a fundamental question for nuclear materials simulation: \textit{can material evolution with atomistic fidelity be advanced across the years-long temporal horizons and whole-RPV spatial regimes required for RPV lifetime prediction?}
As more nuclear reactors worldwide confront long-term service and life-extension decisions, such a capability is of substantial scientific and practical importance~\cite{teller1996completely, poullikkas2013overview, KRIVANEK201499, en10070867, lokhov2012economics}. 
It would enable an unprecedented form of lifetime simulation that directly connects microscopic defect evolution to engineering-scale RPV assessment, thereby supporting multiscale model construction, more reliable lifetime prediction, and better-informed life-extension decisions.

To address this challenge, we present \textit{AtomWorld}, an atomistic world-modeling framework for RPV lifetime simulation co-designed with leadership-scale supercomputing. 
The key insight is that while degradation originates from local atomistic transitions, the phenomena that matter at RPV scales emerge from how those transitions accumulate, interact, and collectively unfold a microscopic material world over long horizons. 
\textit{AtomWorld} therefore models RPV degradation not as a succession of isolated local jumps, but as the unfolding of an atomistic world whose defect state evolves over time toward engineering-relevant material degradation.

Realizing such unfolding at realistic RPV scales requires advancing this microscopic material world across vast temporal and spatial regimes, making large-scale computation and systems co-design essential. \textit{AtomWorld} realizes this shift through three tightly integrated layers:

\textbf{\textit{At the algorithmic level}}, \textit{AtomWorld} recasts classical AKMC as an atomistic world model, where atomic configurations define states and candidate transitions define actions. Instead of advancing evolution through myopic instantaneous-rate sampling, it learns consequence-aware state transitions over the underlying \textit{ab initio} energy landscape. \textit{AtomWorld} combines fixed-radius local atomic policies with strictly $O(1)$ per-atom complexity, a centralized critic that captures long-horizon kinetic structure and distills it into local decisions, and Poisson-based physical time alignment that restores correct temporal semantics under decentralized execution. This formulation establishes a new evolution mechanism that unifies atomistic fidelity, long-horizon kinetic reasoning, and physically consistent scalable time advancement.

\textbf{\textit{At the HPC level}}, \textit{AtomWorld} co-designs this atomistic world model with modern supercomputing architectures, where performance is increasingly constrained by memory movement, synchronization, and communication rather than arithmetic throughput alone. It accordingly derives an execution strategy native to the world-model formulation: compute-dense, massively parallel, and communication-thrifty.  Concretely, \textit{AtomWorld} lifts irregular  transition selection into matrix-centric neural inference through shared-policy aggregation, replaces global-step progression with asynchronous sublattice parallelism, and restructures all-neighbor boundary propagation into dimension-wise shift communication. This co-design yields a compute-dense, synchronization-light, and communication-efficient execution pipeline for large-scale atomistic world modeling.

\textbf{\textit{At the application level}}, \textit{AtomWorld} extends atomistic world modeling to engineering-scale RPV-lifetime simulation of RPV steels through a mesoscopic voxel-parallel framework. Rather than pursuing explicit atom-by-atom reconstruction of the full component, it introduces the voxel as a physically grounded mesoscopic statistical unit that preserves local high-fidelity atomistic evolution while allowing macroscopic material degradation to emerge from the collective behavior of the voxel ensemble.  It further calibrates voxelization through a controlled accuracy–cost tradeoff to preserve local representativeness and kinetically stable defect statistics, and organizes voxel evolution as a heterogeneity-aware task-parallel process under spatially varying irradiation, temperature, and microstructural conditions. This realization elevates atomistic world modeling into a physically grounded and scalable framework for engineering-scale RPV-lifetime simulation.

\section{Current State of the Art}

In this section, we review the major advances in lifetime simulation of RPV steels, from engineering-scale models to explicit atomistic simulation, particularly AKMC.

Historically, progress in RPV lifetime prediction has been driven primarily by engineering-scale modeling, including rate theory~\cite{Odette01061998, KWON20031549}, cluster dynamics~\cite{gokhman2008, KE2022153910, 10.1017/S1431927617000162}, and multiscale coupling~\cite{Odette_Wirth_Bacon_Ghoniem_2001, LIN2024520, CHEIMARIOS20105018, DONG2025124911}.
Rather than explicitly evolving every vacancy jump, solute-defect encounter, cluster nucleation event, or interface transformation, these approaches lift degradation dynamics to coarse-grained kinetic, microstructural, or constitutive variables, allowing direct prediction of macroscopic observables such as defect accumulation, precipitate evolution, hardening, embrittlement, and fracture-relevant property shifts over reactor-relevant domains~\cite{10.1063/1.1839174}.
Their central strength is macroscopic scale reach: they can directly address the spatial and temporal regimes relevant to RPV assessment, spanning vessel-wall thicknesses of hundreds of millimeters, axial dimensions exceeding 10 meters, and service lifetimes of 40--60 years~\cite{Mukhtar_Khattak_Rafique_Zareen_2020, JUMEL2005125, TANAKA201526, knott2013structural}.
This is precisely why they have remained indispensable for engineering design, qualification, and lifetime management. However, this tractability is achieved by replacing explicit atomistic trajectories with fitted evolution laws, effective variables, and constitutive closures.
Under the harsh reactor environments relevant to RPV steels, such abstractions make it difficult to continuously preserve, resolve, or rigorously validate the full microscopic causal chain linking atomistic defect events to long-term degradation~\cite{Odette01061998, gokhman2008, Odette_Wirth_Bacon_Ghoniem_2001, XIE2015196, osti_1111015, osti_1473611}.

To recover the microscopic fidelity lost in these abstractions, subsequent work increasingly turned to explicit atomistic simulation, especially AKMC and related event-driven methods.
Instead of collapsing defect evolution into higher-level phenomenology, these approaches evolve the material directly through discrete atomistic events, including migration, recombination, clustering, emission, and solute-defect interactions~\cite{https://doi.org/10.1002/pssb.200945251
, PhysRevB.90.174102}.
Over the past decade, the HPC community has pushed this line dramatically forward. Starting from synchronous sublattice algorithms~\cite{doi:10.1137/120889459} and continuing through systems such as SPPARKS~\cite{plimpton2009crossing}, Crystal-KMC~\cite{8421558}, OpenKMC~\cite{10.1145/3295500.3356165}, and MISA-AKMC~\cite{10.1145/3712285.3759781}, prior work has expanded the accessible scale of explicit atomistic simulation from millions of atoms to billion-atom systems, and further to trillion- and quadrillion-atom regimes~\cite{10.1145/3295500.3356165, 10.1145/3712285.3759781}.
These advances have been enabled by sparse defect-centric representations, asynchronous domain decomposition, communication-aware parallelization, machine-learning-assisted energetic modeling, and hardware-conscious scheduling~\cite{10.1109/SC41406.2024.00097, xu2023redesigning}. In terms of state size and raw parallel execution, this trajectory is undeniably impressive: it has made explicit atomistic simulation feasible at scales that were previously unimaginable.

Yet these advances remain fundamentally insufficient for RPV-lifetime prediction, because they improve how much atomistic state can be processed in parallel, but not how far the simulation can reach in service-scale physical evolution. In standard KMC, the physical time increment of each step is inversely proportional to the total transition rate. As the simulated volume increases, the candidate event space expands, the aggregate transition rate rises, and the expected physical time advanced per step correspondingly shrinks. Meanwhile, low-barrier and near-reversible local transitions induce severe super-basin trapping, causing the simulation to spend enormous numbers of steps cycling through microscopically valid but macroscopically unproductive events~\cite{10.1145/3295500.3356165, jassar2025challenges}. The practical consequence is stark: even with state-of-the-art AKMC systems on leading supercomputers, explicit simulation would still require on the order of 30 years of wall-clock time to advance just a single service year of RPV material evolution, while record-scale simulations at quadrillion-atom scale still cover only on the order of $10^5 \mu m^3$ of material~\cite{xu2023redesigning, 10.1145/3295500.3356165, 10.1145/3712285.3759781}. This remains negligible compared with the macroscopic regimes required in practice, including vessel-wall thicknesses of hundreds of millimeters and axial dimensions exceeding $10$ meters. In other words, prior AKMC systems have substantially improved state scalability and execution scalability, but they have not resolved physical-time scalability or engineering-scale spatial reach. Machine-learning-assisted atomistic methods can reduce the cost of evaluating local energetics, but as long as they remain within the conventional KMC evolution law, they still inherit the same rate-limited physical-time advancement mechanism~\cite{10.1109/SC41406.2024.00097, 9355242, 10.1145/3581784.3627041, 10.1145/3712285.3771785}.

Overall, existing approaches improve either engineering-scale reach or explicit atomistic fidelity, but not both simultaneously. Engineering-scale models achieve lifetime reach by sacrificing explicit atomistic trajectories, whereas AKMC-style methods preserve microscopic causality but lose physical-time reach as system size grows. The central challenge, therefore, is to make engineering- and lifetime-scale evolution computationally reachable without giving up explicit atomistic fidelity.

\section{Innovations Realized}

\textit{AtomWorld} bridges engineering-scale spatiotemporal scalability and AKMC-level atomistic fidelity, enabling lifecycle simulation of RPV steels with atomic-level accuracy across year-scale temporal horizons and meter-scale spatial dimensions, beyond the reach of traditional approaches.
These capabilities are made possible by the tight integration of AI algorithms—particularly deep reinforcement learning—into the evolution process, together with advances in high-performance computing (HPC) and application-driven innovations.
Fig.~\ref{fig:design} summarizes the key innovations of \textit{AtomWorld} across the algorithmic, HPC, and application levels, with technical details presented in the following sections.

\subsection{Algorithmic Innovation}
\label{design-1}

\begin{figure*}
    \centering
    \includegraphics[width=1\linewidth]{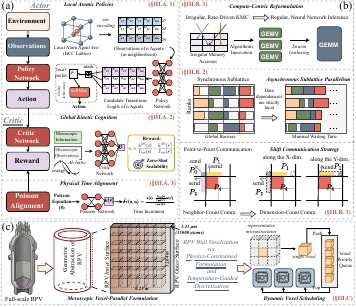}
    \caption{Overview of the key innovations in \textit{AtomWorld}: (a) algorithmic innovation, (b) HPC innovation, and (c) application innovation.}
    \label{fig:design}
\end{figure*}

Instead of advancing dynamics purely through instantaneous rate-driven event sampling, \textit{AtomWorld} represents system evolution as the coordinated behavior of a swarm of physically constrained atomic agents operating over the underlying \textit{ab initio} energy landscape.
Formally, this process is modeled as a Markov decision process $(\mathcal S,\mathcal A,P)$, where atomic configurations correspond to states $\mathcal S$ and candidate atomic transitions define actions $\mathcal A$. 
On this foundation, \textit{AtomWorld} adopts a deep reinforcement learning formulation trained with proximal policy optimization (PPO) within a unified actor--critic framework.
The critic learns the long-horizon kinetic structure of the \textit{ab initio} energy landscape during training, while decentralized policies carry out strictly local decision making at runtime.

This design decomposes \textit{AtomWorld} into three complementary components, as illustrated in Fig~\ref{fig:design}(a):
(1) \textit{Local Atomic Policies} generate candidate transitions from fixed-radius observations, enabling decentralized inference with constant per-atom complexity.
(2) \textit{Global Kinetic Cognition} employs a centralized critic to learn long-horizon kinetic structure and distill it into local policies.
(3) \textit{Physical Time Alignment} reconstructs event time through a Poisson-based formulation, restoring correct AKMC temporal semantics.
Together, these components transform AKMC from a stochastic event sampler into a scalable atomistic world model that advances systems over long physical timescales while preserving strict locality.

\subsubsection{Local Atomic Policies}

\textit{AtomWorld} executes evolution through a swarm of local atom agents. Each atom observes only a finite-radius neighborhood $\mathcal N_i$ and encodes its local configuration as a discrete vector
$
o_i=[\sigma_{ij}]_{j\in\mathcal N_i},
$
where $\sigma_{ij}$ denotes the species label of neighbor $j$. 
Given $o_i$, a shared policy network produces logits over candidate transitions for atom $i$.
To enforce physical feasibility, logits are masked and scaled by a temperature $\tau > 0$:
\begin{equation}
\small
\hat z_{i,k}=
\begin{cases}
z_{i,k}/\tau, & m_{i,k}=1, \\
-\infty, & m_{i,k}=0,
\end{cases}
\end{equation}
where $m_{i,k}$ indicates whether candidate transition $k$ is physically admissible for atom $i$.
Since both the neighborhood size and the action transition $K$ are constant, all atoms evaluate candidate moves independently and in parallel, yielding strictly $O(1)$ per-atom complexity regardless of system size or defect density.
The system-wide action distribution is obtained by concatenating the feasibility-masked logits from all agents and applying a global softmax,
\begin{equation}
\small
    p_\theta(a\mid o_{1:N})=\mathrm{softmax}(\mathrm{concat}(\hat z_1,\ldots,\hat z_N)),
\end{equation}
so that event selection is determined by system-wide competition rather than isolated local decisions.
This differentiable arbitration preserves the global competition mechanism of KMC while enabling end-to-end training.



\subsubsection{Global Kinetic Cognition}

Classical AKMC lacks long-horizon kinetic awareness: decisions are made from instantaneous local rates and therefore cannot distinguish short-lived recrossings from transitions that truly advance structural evolution. 
To address this challenge, \textit{AtomWorld} introduces global kinetic cognition, a centralized training mechanism that teaches the local policies the long-horizon kinetic structure of the \textit{ab initio} energy landscape. 
A global critic aggregates both microscopic agent observations $o_{1:N}$, the \textit{ab initio} energy and mesoscopic descriptors which summarize clustering, vacancy distribution and spatial organization. 
This multi-scale representation allows the critic to infer how repeated local motifs contribute to global basin structure and long-timescale evolution dynamics.
To align learning with physical kinetics, the critic uses a reward derived from the Poisson time potential in \S~\ref{design-1-3}. For each transition $(s, a \rightarrow s')$, the reward is:
\begin{equation}
\small
    r_t=
\frac{\hat u^{(R)}(s)}{\Gamma_{tot}(s)}
-
\frac{\hat u^{(R)}(s')}{\Gamma_{tot}(s')}.
\end{equation}
This reward measures effective physical-time advancement rather than mere configurational change, thereby favoring kinetically meaningful evolution pathways. 

A policy trained on small systems transfers unchanged to larger ones because the global selection probability factorizes over local contexts and depends only on context frequencies and local logits, not on the total system size. Specifically,
\begin{equation}
\small
    \Pr_{\theta}(u,k \mid s_t)=
\frac{\nu_{s_t}(u)\exp\big(z_{\theta}(u)_k\big)}
{\sum_{v\in\mathcal{U}}\nu_{s_t}(v)\sum_{\ell}\exp\big(z_{\theta}(v){\ell}\big)},
\end{equation}
where \(\nu_{s_t}(u)\) is the frequency of local context \(u\); thus, once the local ranking \(u \mapsto z_{\theta}(u)_k\) is learned, the same policy generalizes directly across system sizes with \textbf{zero-shot scalability}.

\subsubsection{{Physical Time Alignment}}
\label{design-1-3}

To reconstruct physically consistent time semantics in \textit{AtomWorld}, we define physical time through the mean first-passage time (MFPT) $\tau(s)$ to an absorbing set~\cite{cai1994statistics, oppelstrup2009first}, which satisfies the Poisson equation by Dynkin's formula
\begin{equation}
\small
\sum_{a\in \mathcal{A}(s)} \Gamma_a(s)\bigl[\tau(\Phi(s,a))-\tau(s)\bigr] + 1 = 0,
\end{equation}
and induces the event-time increment
\begin{equation}
\small
\delta\tau(s,a)=\tau(s)-\tau(\Phi(s,a)),
\end{equation}
where $\mathcal{A}(s)$ denotes the set of feasible events at state $s$.
Introducing the dimensionless potential $u(s)=\Gamma_{\text{tot}}(s)\tau(s)$, we show that under finite-range updates, $u$ admits an exponentially local representation, enabling patch-based approximation with error $O(e^{-\alpha R})$. 
Based on this locality, we train a Poisson Network to predict $u(s)$ from fixed-radius local patches by minimizing a twisted Bellman residual, yielding
\begin{equation}
\small
\hat{\delta\tau}(s,a)=
\frac{
u(s)-\frac{\Gamma_{\text{tot}}(s)}{\Gamma_{\text{tot}}(s')}u(s')
}{
\Gamma_{\text{tot}}(s)
}.
\end{equation}
This formulation reconstructs AKMC-consistent physical-time advancement while preserving decentralized execution, with strictly \(O(1)\) local inference per event.

\begin{figure}
    \centering
    \includegraphics[width=1\linewidth]{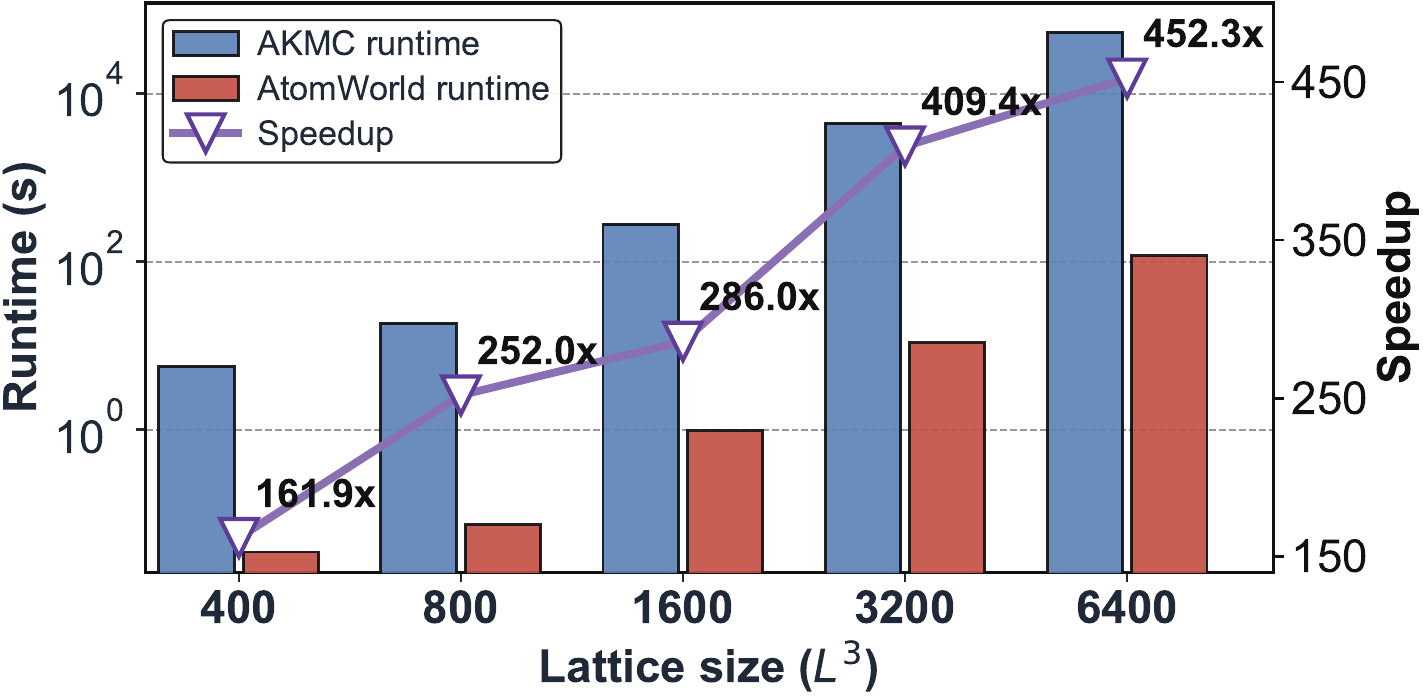}
    \caption{Runtime required to advance one second of physical time for classical AKMC and \textit{AtomWorld} at different lattice sizes, together with the corresponding speedup of \textit{AtomWorld} on a single NVIDIA A100 GPU.}
    \label{fig:lattice_size_speedup}
    \vspace{-3mm}
\end{figure}

Together, these algorithmic innovations fundamentally enhance the efficiency of atomistic evolution, enabling much larger advances in physical time at substantially lower computational cost without sacrificing atomistic fidelity. As shown in Fig.~\ref{fig:lattice_size_speedup}, \textit{AtomWorld} consistently requires far less runtime than classical AKMC to simulate one second of physical time, and its advantage grows with lattice size $L$. The speedup increases monotonically from $161.9\times$ at $L=400$ to $252.0\times$, $286.0\times$, $409.4\times$, and $452.3\times$ at $L=800$, $1600$, $3200$, and $6400$, respectively, demonstrating both strong efficiency and superior scalability. 

\subsection{HPC Innovation}
\label{design-2}

While \textit{AtomWorld} establishes a scalable atomistic world model at the algorithmic level, realizing its full potential requires efficient execution on modern supercomputing architectures. Achieving high efficiency at scale requires not only fast computation, but also a system design that increases compute density, exposes large-scale parallelism, and minimizes communication overhead. As shown in Fig~\ref{fig:design}(b), \textit{AtomWorld} incorporates a set of HPC optimizations that systematically (i) increase arithmetic intensity, (ii) expose large-scale parallelism, and (iii) minimize communication overhead.

\subsubsection{Compute-Centric Reformulation}

In classical AKMC, event selection requires enumerating and evaluating transition rates for all possible migration directions of each vacancy. This procedure is highly irregular and dominated by memory accesses, leaving little room for batching or vectorization. As a result, performance is fundamentally constrained by memory bandwidth, which makes efficient scaling on modern supercomputers difficult.

\textit{AtomWorld} addresses this bottleneck through \textit{compute-centric reformulation}. 
As described in algorithmic innovation (\S~\ref{design-1}), \textit{AtomWorld} replaces explicit transition-rate enumeration with neural network inference for event selection.
This change does more than accelerate physical-time advancement: it converts event selection from an irregular, memory-bound procedure into a regular dense linear algebra workload.
Specifically, each local atom agent performs inference using a shared policy neural network, whose core computation reduces to GEMV operators. 
To further raise computational efficiency, we introduce a \textit{swarm gathering mechanism} that aggregates a large number of independent, weight-sharing GEMV operations into large-scale GEMM kernels. 
This design aligns \textit{AtomWorld} with the matrix-centric execution model of modern supercomputers equipped with dedicated matrix acceleration units. 
Building on this reformulation, \textit{AtomWorld} further improves execution efficiency through mixed-precision computation and kernel fusion. The matrix multiplication components are executed in FP32, which offers substantially higher throughput than FP64 on matrix accelerators while preserving sufficient numerical fidelity for policy inference. 
We also fuse the inference pipeline across network layers to reduce intermediate memory traffic and improve computational throughput.
Together, these optimizations turn event selection from a memory-bound bottleneck into a compute-centric kernel that can efficiently exploit modern supercomputing architectures and scale with future hardware evolution.

\subsubsection{Asynchronous Sublattice Parallelism}
\label{design-2-2}


Conventional synchronous sublattice KMC suffers from severe load imbalance at scale because each evolution superstep ends with a global synchronization, forcing ranks in low-activity regions to wait for those in high-activity regions and thereby leaving many ranks idle and sharply limiting scalability.

\textit{AtomWorld} removes this bottleneck through an \textit{asynchronous sublattice parallelism scheme}. 
Rather than synchronizing globally after each sublattice evolution superstep, each rank advances as soon as the required local dependencies are satisfied.
The key insight is that KMC data dependencies are strictly local: before event selection, a sublattice needs ghost-data synchronization only from its immediate neighboring sublattices.
Under a three-dimensional decomposition, these dependencies are confined to adjacent sublattices along the three spatial directions, while all other state remains local to the same rank. As a result, global synchronization is substantially stronger than what physical correctness and data consistency actually require.
Based on this observation, \textit{AtomWorld} replaces global barriers with a lightweight local dependency mechanism. 
Each sublattice maintains a minimal readiness signal that records only whether the required ghost data from its neighboring sublattices are available. Once these dependencies are met, the sublattice immediately proceeds to event selection and evolution, without waiting for unrelated sublattices or distant ranks.
This design preserves physical correctness and causal consistency while eliminating unnecessary synchronization overhead. 

\subsubsection{Shift Communication Strategy}

As system scale increases, each subdomain must exchange boundary data with all neighboring subdomains simultaneously, causing the number of communication messages to grow rapidly. 
As a result, communication overhead increasingly dominates execution time and becomes a primary scalability bottleneck.

\textit{AtomWorld} alleviates this bottleneck through a \textit{shift communication strategy}. Rather than exchanging boundary data with all neighbors simultaneously, it reorganizes boundary propagation into a dimension-wise pipeline across spatial coordinates. Specifically, the original all-neighbor exchange is decomposed into three sequential stages along the X, Y, and Z axes. At each stage, a subdomain communicates only with its two immediate neighbors along the current dimension and incrementally merges the received boundary states into a local cache for subsequent propagation.
For a three-dimensional decomposition, boundary states are first propagated along the X axis to form an X-extended boundary view, then along the Y axis to construct a two-dimensional neighborhood boundary, and finally along the Z axis to recover the complete three-dimensional boundary information. After these three stages, each subdomain obtains boundary data semantically equivalent to those produced by a direct all-neighbor exchange.
By transforming neighbor-count communication into dimension-count communication, shift communication substantially reduces message count and communication overhead while preserving synchronization semantics and causal consistency.

\subsection{Application Innovation}
\label{design-3}

At the application level, \textit{AtomWorld} extends atomistic world modeling to engineering-scale simulation of RPV steels through a \textit{mesoscopic voxel-parallel framework}. 

\subsubsection{Mesoscopic Voxel-Parallel Formulation}

Rather than reconstructing the entire RPV atom by atom, \textit{AtomWorld} represents RPV-scale degradation as an ensemble of mesoscopic voxel simulations. Each voxel is evolved independently under atomistic dynamics, with no inter-voxel communication during simulation. This yields an embarrassingly parallel formulation and provides a \textbf{zero-communication scaling} path at the application level. Within each voxel, \textit{AtomWorld} resolves local defect generation, migration, clustering, and rare-event kinetics under voxel-specific thermodynamic and irradiation conditions. RPV-scale degradation is then recovered statistically from the aggregated evolution of the voxel ensemble.

\paragraph{Physics-Constrained Formulation}

This decomposition is physically well justified. In engineering practice, RPV steels are modeled using periodic boundary conditions (PBCs) and statistically representative microstructures, rather than explicit full-vessel atomistic reconstruction. The relevant requirement is therefore not global atomistic completeness, but accurate local degradation kinetics under the local temperature, composition, and irradiation conditions. Each voxel thus serves as a \textbf{representative mesoscopic kinetic unit} evolved independently under PBCs.
The voxel size is selected to exceed the characteristic transport--reaction length of mobile defects, so that the dominant migration, interaction, and clustering processes remain self-contained within each voxel. Smaller voxels would suffer from finite-size effects and artificial periodic correlations, whereas larger voxels would mainly increase cost without changing the governing local kinetics. In irradiated Fe-based alloys, this characteristic scale is commonly associated with the inverse sink-strength scale, \(\ell \sim k^{-1}\), and is typically on the order of nanometers to sub-100 nanometers; Cu-rich precipitates are likewise only a few nanometers in size. We therefore use \(2.5\,\mu\mathrm{m}\) mesoscopic voxels, safely above the relevant local kinetic scales by more than one order of magnitude, and recover RPV-scale heterogeneity through massively parallel voxel ensembles rather than larger individual atomistic domains.

\paragraph{Temperature-Guided Discretization}
\label{design-3-1-2}

At the RPV scale, voxels are uniformly discretized along the wall-thickness and axial directions according to the temperature field, as shown in Fig.~\ref{fig:design}(c), with no further discretization along the circumferential direction due to the approximate homogeneity of service conditions in that dimension. Temperature is taken as the primary heterogeneity coordinate because local defect kinetics depend exponentially on it through Arrhenius behavior,
\begin{equation}
\small
r(T)=r_0\exp\!\left(-\frac{E}{k_B T}\right),
\end{equation}
where \(r(T)\) is the local kinetic rate, \(r_0\) is the prefactor, \(E\) is an effective activation barrier, \(k_B\) is the Boltzmann constant, and \(T\) is the absolute temperature. For a small intra-voxel temperature variation \(\Delta T\), the induced relative rate variation satisfies
\begin{equation}
\small
\frac{\Delta r}{r}\approx \Delta \ln r \approx \frac{E}{k_B T^2}\Delta T .
\end{equation}
We therefore choose the voxel count in each direction such that the intra-voxel temperature variation remains below a prescribed tolerance, ensuring that the corresponding variation in local kinetic rates is small. This allows each voxel to be treated as approximately isothermal, while preserving the RPV-scale thermal heterogeneity across the full voxel ensemble.

\subsubsection{Dynamic Voxel Scheduling}

While voxelization makes engineering-scale simulation decomposable, achieving high parallel efficiency is still challenging because voxel runtimes are highly heterogeneous. Differences in temperature, irradiation dose, composition, and defect evolution induce substantial variation in kinetic activity across voxels, making static workload assignment inefficient.
We therefore execute voxel simulations through a \textbf{dynamically scheduled priority queue}. For each voxel \(v\), we compute a lightweight workload proxy
\begin{equation}
\small
    W_v \propto \hat{M}_v \exp\!\left(-\frac{\hat{E}_v}{k_B T_v}\right),
\end{equation}
where \(T_v\) is the voxel temperature, \(\hat{M}_v\) characterizes local event multiplicity, and \(\hat{E}_v\) is an effective activation barrier determined by the current defect state. Larger \(W_v\) indicates higher expected kinetic intensity and, correspondingly, a heavier computational task.
At runtime, voxels with larger \(W_v\) are dispatched earlier, and each node pulls a new voxel immediately after finishing its current one. This online scheduling policy reduces tail imbalance from heterogeneous voxel runtimes and turns static spatial decomposition into adaptive task parallelism, maintaining high utilization and strong scalability.

Together, mesoscopic voxelization and dynamic scheduling bridge atomistic fidelity and engineering-scale simulation. \textit{AtomWorld} resolves local kinetics within each voxel, and large-scale parallel aggregation lifts them into RPV-scale estimates of lifetime degradation. Without explicit full-component atomistic reconstruction, this framework enables predictive RPV simulation with controlled accuracy, practical cost, and strong scalability.


\section{How Performance Was Measured}

\subsection{HPC systems}

We evaluate our method on four supercomputers, which we group into two categories according to their architectural characteristics: CPU-based supercomputers optimized for large-scale general-purpose scientific computing, and GPU-based supercomputers designed around accelerator-centric high-throughput computation.

\textbf{{\textit{CPU-based Supercomputers:}}}



\textit{Tianhe-3} consists of more than 110,000 nodes with a theoretical FP64 peak of 1.6~EFLOPS. Each node is built around the MT-3000 heterogeneous processor, which delivers up to 11.5~TFLOPS and integrates 16 general-purpose cores, 96 control cores, and 1,536 accelerator cores organized into one GP zone and four acceleration zones. Each acceleration zone is equipped with 48~MB HBSM and 32~GB DDR4 memory. 

\textit{New Sunway} contains 107,520 nodes and achieves a theoretical FP64 peak performance of 1.5~EFLOPS. Each node is powered by a single SW26010 Pro processor with 96~GB memory, partitioned across six core groups. Each core group contains one management processing element and 64 compute processing elements arranged in an $8 \times 8$ mesh. All nodes are connected through a fat-tree network to support large-scale parallel execution.

\textbf{{\textit{GPU-based Supercomputers:}}}

{\textit{ORISE}, a {\textit{leading}} GPU-based supercomputer in China, consists of 7,086 nodes and provides a theoretical peak performance of 200~PFLOPS.}
Each node is equipped with one 32-core Hygon C86 7185 CPU, organized into four 8-core NUMA domains, and four Hygon DCU accelerators. The DCUs are HIP-based GPGPUs comparable to AMD MI60-class devices, each with 16~GB of dedicated VRAM. Each node also provides 128~GB of host memory. 
Within a node, CPU cores communicate through the HSL high-speed interconnect bus protocol, while CPU--accelerator communication is performed via PCIe-based DMA.
Inter-node communication is provided by a 200~Gb/s network.

\textit{Tecorigin} consists of 512 liquid-cooled T1118L nodes and delivers a theoretical peak performance of 1.31~EFLOPS in FP16. Each node includes two 32-core Loongson 3C6000/D processors and eight proprietary T111 OAM accelerator modules based on a heterogeneous many-core design. 
Nodes are interconnected through a two-level fat-tree InfiniBand network for high-throughput distributed execution.

\subsection{Physical system used to measure performance}

Under the simplified engineering approximation adopted in this work, the local service conditions of voxel \(v\) are determined by both its through-wall position \(x_v\) and its axial position \(z_v\) in the Chinese third-generation CAP1400 RPV~\cite{yan2024experimental}. The RPV base material is ASME SA508 Grade 3 Class 1, and we adopt a representative composition reported for China domestic A508-3 steel: Fe (bal.), C 0.167 wt.\%, Si 0.193 wt.\%, Mn 1.35 wt.\%, S 0.002 wt.\%, P 0.005 wt.\%, Cr 0.086 wt.\%, Ni 0.738 wt.\%, Cu 0.027 wt.\%, Mo 0.481 wt.\%, and V 0.007 wt.\%~\cite{Ma_2021}. 

The local irradiation condition is prescribed as:
\begin{equation}
\small
\phi_v = \phi_{\mathrm{inner}} \exp(-\mu x_v)\, f_\phi(z_v),
\end{equation}
where \(\phi_{\mathrm{inner}}\) is the reference neutron flux at the inner wall, \(\mu\) is the through-wall attenuation coefficient, and \(f_\phi(z_v)\) describes the axial flux distribution, which peaks in the core belt region, as illustrated in Fig.~\ref{fig:overall}(b).
Accordingly, the initial vacancy concentration in voxel \(v\) is treated as a function of its local service conditions:
\begin{equation}
\small
c_{V,v}^{(0)} = c_V^{(0)}\!\left(T_v,\phi_v,c_{\mathrm{solute},v},\rho_{\mathrm{sinks},v}\right),
\end{equation}
where \(c_{\mathrm{solute},v}\) denotes the local solute composition and \(\rho_{\mathrm{sinks},v}\) the effective density of defect sinks such as dislocations, grain boundaries, and precipitates. 
In this way, the initial defect state of each voxel reflects the engineering-scale spatial heterogeneity, while the subsequent evolution of vacancy concentration and related defect statistics is governed self-consistently by \textit{AtomWorld}.

With this formulation, engineering-scale degradation in CAP1400 RPV steel is recast as a massively parallel dynamical evolution problem over a large ensemble of local atomistic boxes, thereby achieving both local physical fidelity and system-level scalability.

\subsection{Training details}


All training of the algorithm part was conducted on single NVIDIA A100 GPUs using PyTorch 2.5.1 with CUDA 12.4. 
The model was trained directly in an AKMC simulation environment on systems of size \(200 \times 200 \times 200\). 
It exhibits \emph{zero-shot system-size scalability}: once trained on small lattices, it can be directly deployed to much larger systems without retraining.
The training data were constructed from atomistic transition trajectories sampled across a broad range of material and environmental conditions, enabling the learned model to generalize beyond any single fixed system. 
Specifically, the sampled trajectories cover temperatures from 230 to 400 $^\circ\mathrm{C}$, alloy compositions spanning Cu = 0.02--0.26~at.\%, Ni = 0.38--1.54~at.\%, Mn = 0.59--1.57~at.\%, Si = 0.10--0.99~at.\%, and P = 0.004--0.041~at.\%, point-defect concentrations from 1 to 1000~appm, neutron fluxes from \(10^{9}\) to \(10^{11}\,\mathrm{n\,cm^{-2}\,s^{-1}}\), and accumulated irradiation doses from \(10^{-4}\) to 1~dpa. Each sample consists of a global state \(s_t\), the local contexts of all active atoms, the corresponding candidate transition sets, and the supervision targets for the policy, value, and time branches. For each active atom, we extract a fixed-radius local neighborhood with cutoff radius \(6.0~\AA\), and cap the maximum number of neighbors at 64; excess neighbors are truncated and smaller neighborhoods are zero-padded with masking. The input features include atom type, relative coordinates, local defect type, neighborhood connectivity, and candidate-transition masks.

\textit{AtomWorld} consists of a local atomic policy network, a global kinetic critic, and a Poisson time network. The local policy network takes the local context of a single active atom as input and outputs logits over all valid candidate transitions, while the critic and time branches provide global kinetic supervision and physical-time prediction during training. The model is trained with a joint objective combining a policy loss, a critic regression loss, and a time-alignment loss, optimized using AdamW with batch size 256 and initial learning rate \(10^{-4}\). The final model is selected based on the best validation performance.

At simulation time, only the local policy network and the Poisson time network are retained, while the global critic is used only during centralized training. Since inference depends only on fixed-radius local neighborhoods, the per-atom inference cost remains constant with respect to system size, and models trained on small systems can be directly applied to larger systems without changing the parameters or network architecture.

\subsection{Measurement methodology}

The proposed method will be evaluated along four dimensions.
\textit{First}, correctness will be validated by comparing our results against the experimental measurements reported by Lê et al. and the simulation results presented by Vincent et al. This ensures that the method remains faithful to both physical observations and established computational references.
\textit{Second}, scalability will be assessed on supercomputers with diverse architectural designs. This evaluation is intended to verify that the method can sustain efficient execution across heterogeneous large-scale platforms.
\textit{Third}, computational performance will be quantified using a conservative FLOP-based methodology. The total FLOP count is derived from the exact arithmetic complexity of each kernel, accumulated locally on each MPI rank, and reduced globally at the end of execution, so the reported performance reflects effective scientific computation rather than inflated hardware activity.
\textit{Fourth}, for full-size RPV lifetime prediction, we use time-to-solution as the primary metric, 
defined as the execution time required to advance one year of physical service time.
This metric more directly captures the practical cost of full-scale lifecycle simulation.

\section{Performance Results}

\begin{figure}[t]
    \centering
    \includegraphics[width=1\linewidth]{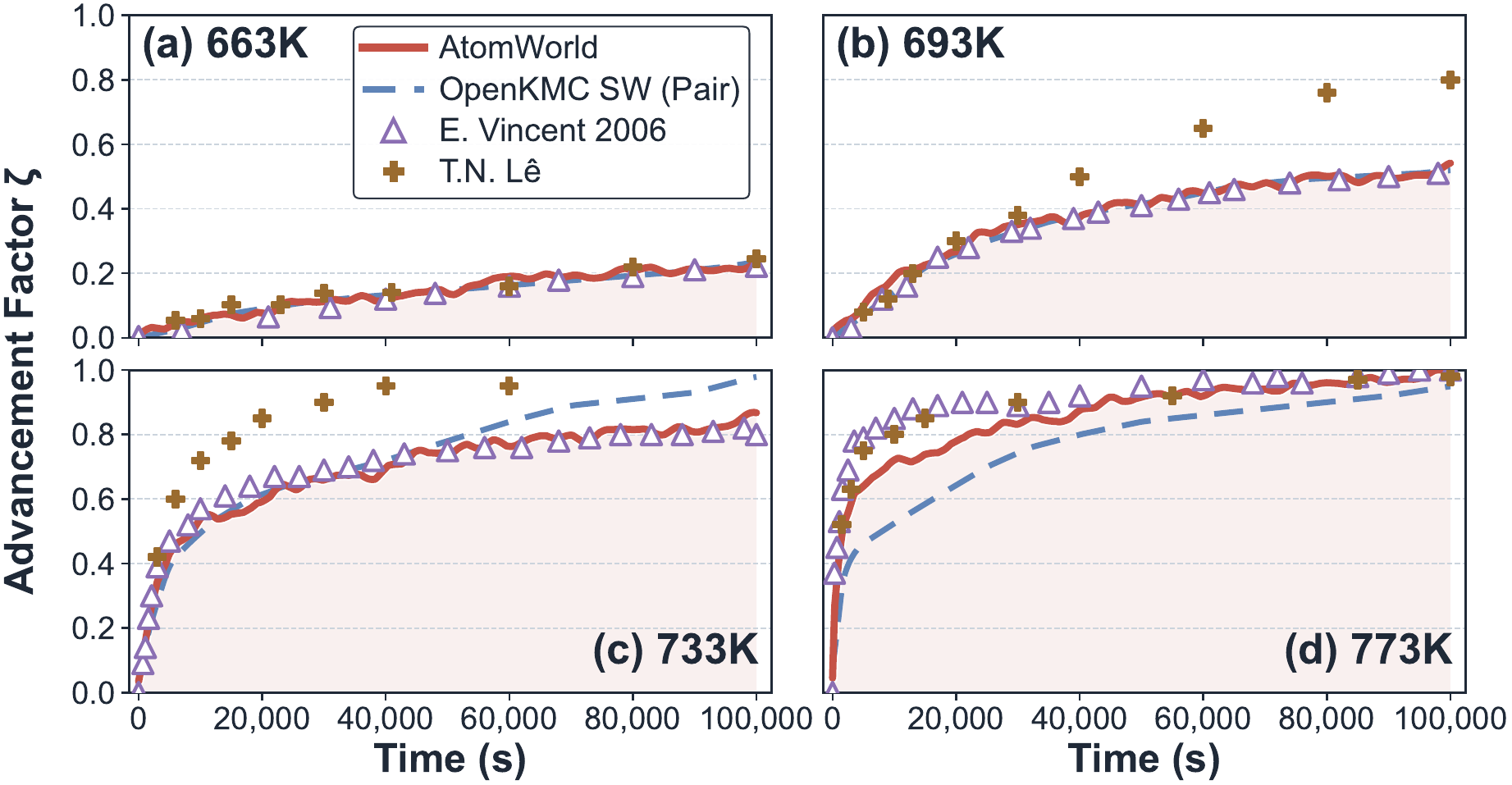}
    \caption{Time evolution of the advancement factor $\zeta$ at different temperatures. \textit{AtomWorld} closely reproduces the reference KMC trajectories across thermal regimes, capturing both slow low-temperature evolution and faster high-temperature kinetics.}
    \label{fig:zeta_vs_time_atomworld}
\end{figure}

\begin{figure*}[t]
    \centering
    \includegraphics[width=1\linewidth]{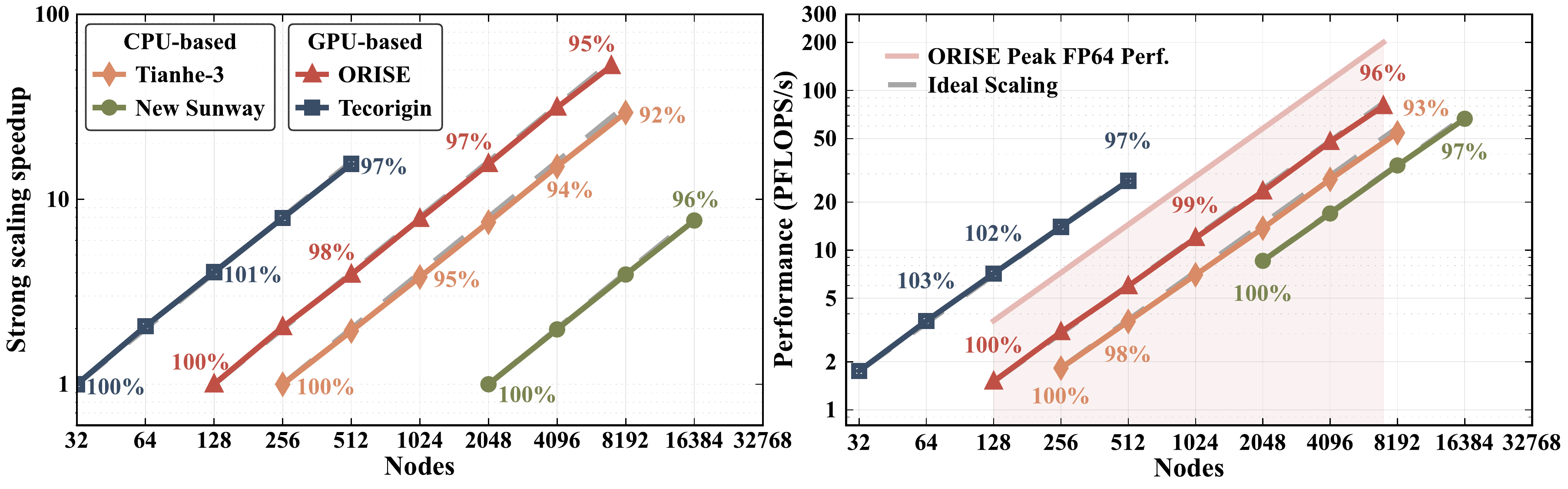}
    \caption{
    Strong scalability (left) and weak scalability (right) results.
    Numbers along the graph lines indicate parallel efficiency.
    Tecorigin performance is measured with FP16 kernels, while the other platforms use mixed FP32/FP16 kernels.
    }
    \label{fig:scaling}
    \vspace{-3mm}
\end{figure*}

\subsection{Accuracy Validation}

A central question is whether \textit{AtomWorld} can accurately capture the physical evolution of the underlying atomistic system.
To assess this, we examine the advancement factor $\zeta(t)$, which tracks microstructural progression over physical time across different thermal conditions.

As shown in Fig.~\ref{fig:zeta_vs_time_atomworld}, \textit{AtomWorld} closely matches the reference AKMC trajectories across all tested temperatures. It captures both the slow evolution at low temperature and the faster progression at high temperature, without distorting the time profile. Similar agreement is observed in the corresponding energy-relaxation trajectories, indicating that \textit{AtomWorld} preserves both configurational and thermodynamic evolution. Together, these results show that \textit{AtomWorld} accelerates atomistic simulation without sacrificing the essential physics of the original kinetics.

\subsection{Scalability Evaluation}

\begin{table}[b]
\centering
\caption{Machine-specific voxel configurations across different supercomputers.}
\label{tab:voxel_size}

\footnotesize
\begin{tabular}{lccc}
\toprule
\textbf{\textit{Supercomputer}} & \textbf{Voxel size} & \textbf{\#Atoms} & \textbf{Volume} \\
\midrule


\textit{Tianhe-3}  & $7{,}500^{3}$  & $844\,B$  & $9.93\,\mu m^3$ \\

\textit{New Sunway} & $10{,}000^{3}$ & $2\,T$    & $23.54\,\mu m^3$ \\

\textit{ORISE}  & $10{,}000^{3}$ & $2\,T$    & $23.54\,\mu m^3$ \\

\textit{Tecorigin} & $20{,}000^{3}$ & $16\,T$   & $188.33\,\mu m^3$ \\

\bottomrule
\end{tabular}
\end{table}

We evaluate the scalability of \textit{AtomWorld} on four supercomputers.
In all experiments, \textit{AtomWorld} adopts a two-level parallelization strategy: inter-node parallelism is provided by the \textit{voxel-parallel} scheme (\S~\ref{design-3}), while intra-node parallelism is realized through the \textit{asynchronous sublattice-parallel} design (\S~\ref{design-2-2}).
To accommodate architectural differences, we adopt machine-specific voxel configurations for node-local execution, as summarized in Table~\ref{tab:voxel_size}.
We then tailor each system’s scaling setup accordingly, as listed in Table~\ref{tab:scaling_config}, to evaluate distributed execution at hardware-efficient operating points and fully expose the strong- and weak-scaling capability of \textit{AtomWorld}.

As shown in Fig.~\ref{fig:scaling}, \textit{AtomWorld} delivers consistently strong strong- and weak-scaling performance across all four leadership systems. 
On \textit{Tianhe-3} and \textit{New Sunway}, \textit{AtomWorld} likewise maintains high scaling efficiency, achieving 29.4$\times$ and 7.7$\times$ strong-scaling speedups over 32$\times$ and 8$\times$ increases in node count, from 256 to 8,192 nodes and from 2,048 to 16,384 nodes, respectively, with corresponding strong-scaling efficiencies of 92\% and 96\%, and weak-scaling efficiencies of 93\% and 97\%.
Similar trends hold on GPU-based systems: 
on \textit{ORISE}, the most advanced GPU-based supercomputer in China, \textit{AtomWorld} achieves a 52.2$\times$ strong-scaling speedup with 95\% efficiency as the system scales by 55$\times$, from 128 to approximately 7,086 nodes, while sustaining 96\% weak-scaling efficiency; 
on \textit{Tecorigin}, it delivers 15.5$\times$ speedup in both strong and weak scaling with 97\% efficiency over a 16$\times$ increase in machine scale, from 32 to 512 nodes.

\subsection{Peak Performance}

Figure~\ref{fig:scaling} (right) reports the peak performance of \textit{AtomWorld}, measured as the floating-point throughput of the core neural-network inference that dominates the simulation runtime.
On ORISE, \textit{AtomWorld} reaches 80.0~PFLOPS on 7,086 nodes for the system with 708,600 voxels and 2 trillion atoms per voxel, corresponding to 40\% of the system's peak FP64 performance.
Overall, these results show that \textit{AtomWorld} not only scales to supercomputer deployment, but also sustains substantial throughput on the dominant inference workload at extreme scale.

\begin{table}[t]
\centering
\caption{Scaling configurations across different supercomputers.}
\label{tab:scaling_config}

\begin{adjustbox}{width=1\linewidth}
\begin{tabular}{lcccccc}
\toprule

\multirow{2}{*}{\textbf{\textit{Supercomputer}}} & 
\multirow{2}{*}{\textbf{Nodes}} & 
\multicolumn{2}{c}{\textbf{Strong Scaling}} & 
\multicolumn{2}{c}{\textbf{Weak Scaling}} \\
\cmidrule(rl){3-4} \cmidrule(rl){5-6}
&  & \textbf{\#Voxels} & \textbf{Voxels/Node} & \textbf{\#Voxels} & \textbf{Voxels/Node} \\

\midrule



\multirow{2}{*}{\textit{Tianhe-3}}
& 256  & 409{,}600 & 1{,}600 & 128{,}000 & 50 \\
& 8{,}192 & 409{,}600 & 50   & 409{,}600 & 50 \\

\midrule

\multirow{2}{*}{\textit{New Sunway}}
& 2048  & 819200 & 400 & 400    & 50 \\
& 16384 & 819200 & 50  & 819200 & 50 \\

\midrule

\multirow{2}{*}{\textit{ORISE}}
& 128  & 256{,}000 & 2{,}000 & 12{,}800  & 100 \\
& 7{,}086 & 256{,}000 & 36.1 & 708{,}600 & 100 \\

\midrule

\multirow{2}{*}{\textit{Tecorigin}} 
& 32  & 25{,}600 & 800  & 1{,}600   & 50 \\
& 512 & 25{,}600 & 50   & 25{,}600  & 50 \\

\bottomrule

\end{tabular}
\end{adjustbox}

\end{table}

\section{Implications}



\textit{AtomWorld} redefines what is computationally reachable for RPV lifetime simulation.
For the first time, direct atomistic simulation is no longer confined to probing microscopic mechanisms within limited spatial and temporal windows; instead, it can be advanced across service-year horizons and engineering-scale RPV dimensions.
This lifts atomistic simulation from a tool primarily used to interpret local phenomena into a predictive instrument that can directly inform RPV-scale degradation assessment.
As a result, \textit{AtomWorld} establishes a new computational foundation for physically grounded lifetime prediction, cross-scale model development, and more reliable life-extension decisions in safety-critical nuclear energy systems.
It also opens the possibility of using atomistic simulation not only to explain degradation after the fact, but to anticipate it before irreversible risks emerge.

More broadly, this work points to a new route for scientific computing at extreme scale.
Its significance lies not merely in running larger simulations faster, but in showing that previously inaccessible scientific regimes can be unlocked by reformulating the evolution process itself and co-designing it with modern supercomputing.
By recasting atomistic simulation as world modeling, \textit{AtomWorld} demonstrates that long-horizon physical evolution can be made simultaneously atomistically faithful, temporally reachable, spatially extensive, and machine-efficient.
This shifts the central question from how to accelerate isolated kernels to how to restructure an entire physical evolution problem so that scientific fidelity and extreme-scale execution become mutually reinforcing rather than fundamentally conflicting.
In this sense, \textit{AtomWorld} suggests a broader paradigm shift for computational science: future breakthroughs may come not only from larger machines, but from new formulations that transform how complex physical worlds are represented, advanced, and scaled.


\clearpage

\bibliographystyle{IEEEtran}
\bibliography{main}

\end{document}